\documentclass[12pt]{article}
\usepackage{hfstyle}
\usepackage{amscd,amsthm,graphicx,latexsym}
\usepackage{bbm}
\usepackage{amssymb,amsmath}
\usepackage{hyperref}

\newtheorem{theorem}{Theorem}[section]
\newtheorem{lemma}[theorem]{Lemma}

\newcommand{\ZZ}{{\mathbbm{Z}}}
\newcommand{\NN}{{\mathbbm{N}}}

\newcommand{\B}{{\mathcal B}}
\newcommand{\G}{{\mathcal G}}
\newcommand{\calS}{{\mathcal S}}
\newcommand{\f}{{\mathbf f}}

\newcommand{\card}{\mathop{\mathrm{card}}}

\begin{document}
\date{November 7, 2007}
\title{Catalan numbers and power laws in cellular automaton rule 14}

\author{Henryk Fuk\'s\thanks{
        Corresponding author, email {\tt hfuks@brocku.ca}. Author acknowledges financial 
support from the Natural Sciences and Engineering Research Council of Canada (NSERC) in the
form of Discovery Grant.}  \,\,\,and Jeff Haroutunian\thanks{Current address:
University of Calgary, Department of Mathematics and Statistics, Calagary,
Alberta T2N 1N4, Canada.}\\
      \oneaddress{
         Department of Mathematics\\
         Brock University\\
         St. Catharines, ON L2S 3A1, Canada
       }
   }

\Abstract{
We discuss example of an elementary cellular automaton for which the
density of ones decays toward its limiting value as a power of the number
of iterations $n$.  Using the fact
that this rule conserves the number of blocks 10 and that preimages of some
other blocks exhibit patterns closely related to patterns observed in rule 184,
we derive expressions for the number of $n$-step preimages of all blocks of length 3.
These expressions involve Catalan numbers, and together with basic properties
of iterated probability measures they allow us to 
to compute the density of ones  after $n$ iterations, as well as probabilities
of occurrence of arbitrary block of length smaller or equal to 3. 
}
\maketitle
\section{Introduction}
A question of interest in the theory of cellular automata (CA) is the
action of the global function of a given cellular automaton on an
initial probability measure. Typically, one starts with a Bernoulli
measure and wants to determine the measure after $n$ applications of
the CA rule. More informally, we start with an initial configuration
where each site is (in the case of binary rules) equal to $1$ with
probability $p$ and equal to $0$ with probability $1-p$, independently for all
sites. Then we ask: after $n$ iterations of the CA rule, what is the
probability of occurrence of a given block $b$ in the resulting configuration?
The simplest question
of this type is often phrased informally as follows: what is the density
of ones $\rho_n$ after $n$ iterations of the CA rule over a ``random'' initial
configuration with $50\%$ of zeros and $50\%$ of ones?  Here by ``density''
one understands the probability of occurrence of 1.

It is usually very hard to answer questions like this rigorously, and
quite often one has to resort to numerical experiments. Such experiments
reveal that for many cellular automata, $\rho_n$ exponentially decays
toward some limiting value $\rho_\infty$. In physics  this is
sometimes called ``exponential relaxation to equilibrium''. Other types
of relaxation to equilibrium, such as power law decay toward $\rho_\infty$,
are less common in CA, and to our knowledge, no such case has ever been discussed
rigorously in the CA literature.
 
In this paper we demonstrate that in the elementary CA rule 14 the decay
of $\rho_n$ toward its limiting value follows a power law. We show that
the origin of this power law is the scaling of the numbers of $n$-step 
preimages of certain finite blocks. The precise numbers of preimages 
of these blocks can be expressed in terms of Catalan numbers, similarly
as previously reported for the rule 184 \cite{paper11}. For large $n$, one  can approximate
Catalan numbers via Stirling formula, obtaining as a result the power law 
$\rho_n -\rho_\infty \sim n^{-1/2}$.

The paper is organized as follows. Following basic definitions,
we present a theorem on enumeration of preimages of basic blocks 
in rule 14. The poof of this theorem requires four technical lemmas,
which are proved in sections following the proof of the main theorem.
The formula for density of ones is then proved using the enumeration
theorem, and the asymptotic power law is derived.

\section{Basic definitions}
Let $\G=\{0,1,...N-1\}$ be called {\em a symbol set}, and let $\calS(\G)$
 be the set of all bisequences over $\G$, where by a bisequence we mean a
 function on  $\ZZ$ to $\G$. Set  $\calS(\G)$, which
 is a compact, totally disconnected, perfect, metric space,  will be
called {\em the configuration space}. Throughout the remainder of this
text
 we shall  assume that $\G=\{0,1\}$, and  the configuration space
 $\calS(\G)=\{0,1\}^{\ZZ}$ will be simply denoted by $\calS$.

{\em A block of length} $r$ is an ordered set $b_{0} b_{1}
\ldots b_{n-1}$, where $n\in \NN$, $b_i \in \G$.
Let $n\in \NN$ and let
$\B_n$ denote the set of all blocks of length $n$ over $\G$. The
number of elements of $\B_n$ (denoted by $\card \B_r$) equals
$2^{n}$. 

For $r \in \NN$, a mapping $f:\{0,1\}^{2r+1}\mapsto\{0,1\}$ will be called {\em a cellular
 automaton rule of radius $r$}. Alternatively, the function $f$ can be
 considered as a mapping of $\B_{2r+1}$ into $\B_0=\G=\{0,1\}$. 

Corresponding to $f$ (also called {\em a local mapping}) we define a
 {\em global mapping}  $F:\calS \to \calS$ such that
$
(F(s))_i=f(s_{i-r},\ldots,s_i,\ldots,s_{i+r})
$
 for any $s\in \calS$.
The {\em composition of two rules} $f,g$ can be now defined in terms
of
their corresponding global mappings $F$ and $G$ as $
(F\circ G)(s)=F(G(s)),
$
where $s \in \calS$. 

A {\em block evolution operator} corresponding to $f$ is a mapping
 $\f:\B \mapsto \B$ defined as follows. 
Let $r\in \NN$ be the radius of $f$, and let  $a=a_0a_1 \ldots a_{n-1}\in \B_{n}$
where $n \geq 2r+1 >0$. Then 
\begin{equation}
\f(a) = \{ f(a_i,a_{i+1},\ldots,a_{i+2r})\}_{i=0}^{n-2r-1}.
\end{equation}
 Note that if
$b \in B_{2r+1}$ then $f(b)=\f(b)$.

We will consider the case of $\G=\{0,1\}$ and $r=1$ rules,
 i.e., {\em elementary cellular automata}. In this case, when $b\in\B_3$,
then $f(b)=\f(b)$. The set 
$\B_3=\{000,001,010,011,100,101,101,110$, $111\}$ will be called the set of \textit{basic blocks}.

The number of $n$-step preimages of the block $b$ under the rule $f$
is defined as the number of elements of the set $\f^{-n}(b)$.
Given an elementary rule $f$, we will be especially interested in
the number of $n$-step preimages of basic blocks
under the rule $f$.

\section{Rule 14}
Local function of the elementary cellular automaton rule 14 is defined as
\begin{equation} \label{r14def}
f(x_0,x_1,x_2)=x_1+x_2+x_1 x_0 x_2-x_1 x_2-x_0 x_2-x_1 x_0.
\end{equation}
This means that $f(0,0,1)=f(0,1,0)=f(0,1,1)=1$, and $f(x_0,x_1,x_2)=0$
for all other triples $(x_0,x_1,x_2)\in \{0,1\}^3$.

Preimage sets of basic blocks, that is, sets ${\f}^{-n}(b)$ for
$b \in \B_3$, have rather complex structure for this rule.
Figure 1 shows, as an example, preimages ${\f}^{-n}(b)$ for $b=101$ for $n=1,2,3$.
Preimages are represented in the form of a tree rooted at 101. Preimages ${\f}^{-1}(101)$
are 01010, 01001, and 01011, and they are shown as the first level of the tree, ${\f}^{-2}(101)$
as the second level, and ${\f}^{-3}(101)$ as the third level. When one block $a$ is the image of another block
$b$, that is, $\f(b)=a$, an edge from $a$ to $b$ is drawn.

In spite of the apparent complexity of sets ${\f}^{-n}(b)$ for basic blocks $b$, it is possible to determine cardinalities of these sets. 

\begin{figure}[t]
 \centering
 \begin{center}
 \includegraphics[scale=1.2]{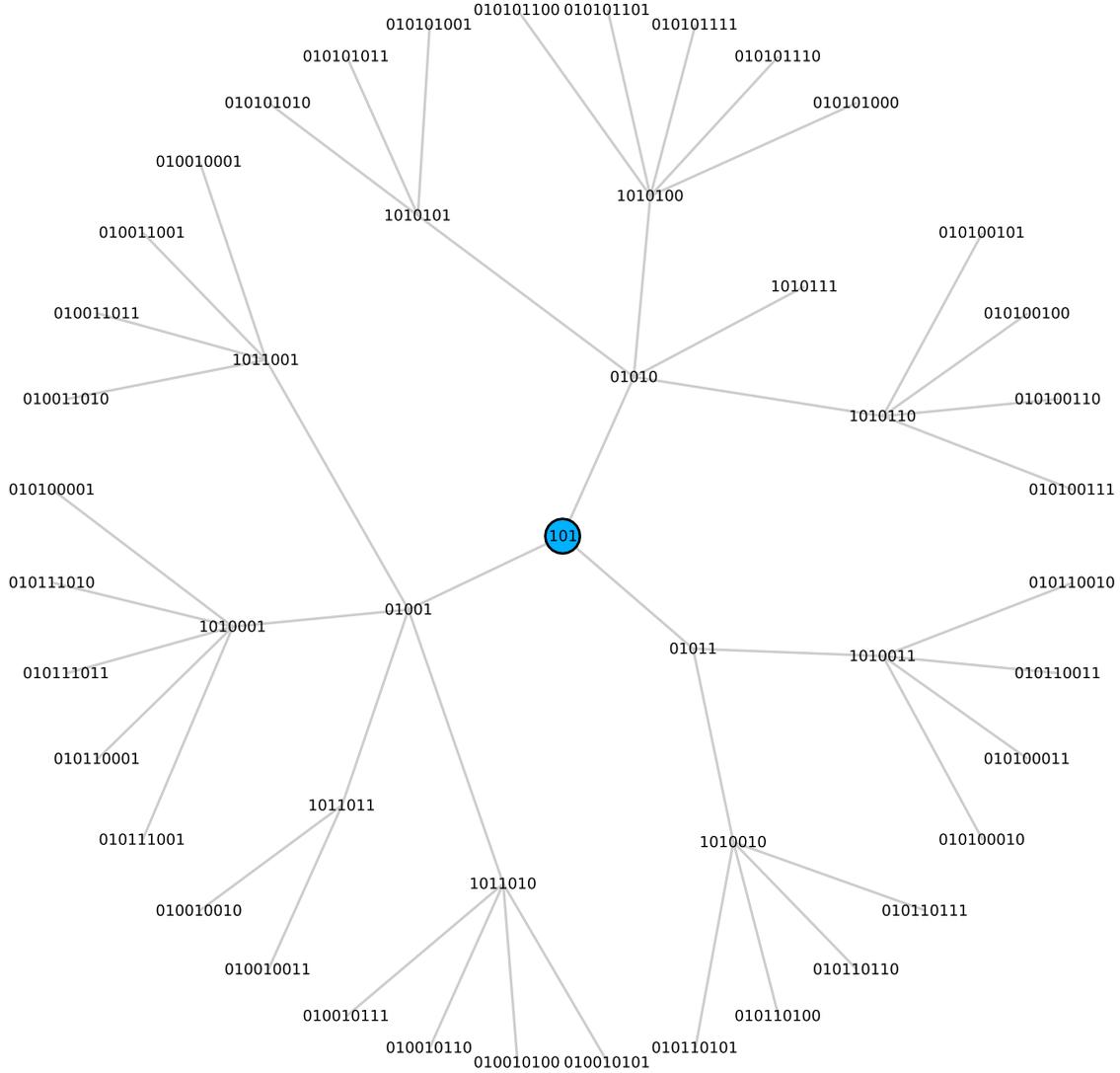}
\end{center}
 \caption{Preimage tree for the block  101 for rule 14.}
\end{figure}

\begin{theorem}\label{maintheorem}
Let $\f$ be the block evolution operator for CA rule 14.
Then for any positive integer $n$ we have
\begin{eqnarray*}
\card \f^{-n}(000)&=&(4n+3)C_n,\\
\card \f^{-n}(001)&=&2^{2n+1}-(2n+1)C_n,\\
\card \f^{-n}(010)&=&2(n+1)C_n,\\
\card \f^{-n}(011)&=&2^{2n+1}-2(n+1)C_n,\\
\card \f^{-n}(100)&=&2^{2n+1}-(2n+1)C_n,\\
\card \f^{-n}(101)&=&(2n+1)C_n,\\
\card \f^{-n}(110)&=&2^{2n+1}-2(n+1)C_n,\\
\card \f^{-n}(111)&=&0,
\end{eqnarray*}
where $C_n$ is the $n$-th Catalan number
\begin{equation}
C_n=\frac{1}{2n+1} \binom{2n}{n}=\frac{(2n)!}{n!(n+1)!}.
\end{equation}
\end{theorem}
Proof of this theorem will be based on the following four lemmas.
\begin{lemma} \label{l0}
Let 
\begin{eqnarray}
f_{184}(x_0,x_1,x_2)&=&x_0+x_1 x_2-x_1 x_0,\\
f_{195}(x_0,x_1,x_2)&=&1-x_1-x_0+2 x_1 x_0
\end{eqnarray}
for $x_0,x_1,x_2 \in \{0,1\}$. Then if  $x_0,x_1,x_2, x_3,x_4 \in \{0,1\}$
and at least of one of $x_1$, $x_2$ is equal to zero, we have
\begin{align}
 f_{195}\big(f_{14}(x_0,x_1,x_2),&f_{14}(x_1,x_2,x_3),f_{14}(x_2,x_3,x_4)\big)=\\ \nonumber
 &f_{184}\big(f_{195}(x_0,x_1,x_2),f_{195}(x_1,x_2,x_3),f_{195}(x_2,x_3,x_4)\big)
\end{align}

\end{lemma}

\begin{lemma} \label{l1}
For any  $n\in \NN$, the number of $n$-step preimages of 101 under the rule 14
is the same as the number of $n$-step preimages of 000 under the rule 184, that is,
\begin{equation}
\card \f^{-n}_{14}(101) = \card \f^{-n}_{184}(000),
\end{equation}
where subscripts 184 and 14 indicate block evolution 
operators
for, respectively, CA rules 184 and 14. Moreover, the bijection $M_n$ from the set $\f^{-n}_{184}(000)$
to the set $\f^{-n}_{14}(101)$ is defined by
\begin{equation}
M_n(x_0x_1\ldots x_m)=\left\{n+j+1+\sum_{i=0}^j x_i\mod 2\right\}_{j=0}^{m}
\end{equation}
for  $m\in \NN$ and for $x_0x_1\ldots x_m \in \{0,1\}^m$.
\end{lemma}

\begin{lemma} \label{l3}
For any $n\in N$, we have
\begin{equation}
\card \f_{14}^{-n}(10)= \card \f_{14}^{-n}(01)=4^n.
\end{equation}
\end{lemma}

\begin{lemma} \label{l4}
For any $n\in N$,
\begin{equation}
\f_{14}^{-n}(101)=R\left[R\left[\f_{14}^{-n-1}(010)\right]\right],
\end{equation}
where R is the right truncation operator defined as $R(b_0b_1 \ldots b_n)=b_0b_1 \ldots b_{n-1}$.
\end{lemma}

\section{Proof of the main theorem}
We will prove the theorem first assuming validity of all four lemmas.  Proofs of lemmas will be presented in section \ref{lemmaproofs}.

Since Lemma~\ref{l1} implies that sets $\f^{-n}_{184}(000)$ and $\f^{-n}_{14}(101)$ 
have the same cardinality, and since it has been shown in \cite{paper11} that
\begin{equation}
\card \f^{-n}_{184}(000) =(2n+1)C_n,
\end{equation}
we immediately obtain the desired result  $\card \f^{-n}_{14}(101) =(2n+1)C_n$.

Lemma~\ref{l4} immediately yields 
\begin{equation}
\card \f^{-n}(010)=2(n+1)C_n.
\end{equation}

Since by definition of $\f$, $\card \f^{-n}(101)+ \card \f^{-n}(100)=2\card \f^{-n}(10)$,
we obtain
\begin{equation}
 \card \f^{-n}(100)=2\card \f^{-n}(10)-\f^{-n}(101),
\end{equation}
and using Lemma~\ref{l3},
\begin{equation}
\card \f^{-n}(100)=2^{2n+1}-(2n+1)C_n.
\end{equation}
Similar argument leads to
\begin{equation}
\card \f^{-n}(001)=2^{2n+1}-(2n+1)C_n.
\end{equation}

The formula for $\card \f^{-n}(011)$ can be obtained by observing, again
from the definition of $\f$, that $\card \f^{-n}(011)+ \card \f^{-n}(010)=2\card \f^{-n}(01)$, and by using Lemma~\ref{l3}. Proof of the formula for $\card \f^{-n}(110)$
is similar.

Thus, the only cases which we are missing are 000 and 111. The case of
111 is a direct consequence of the fact that the block 111 has no preimages 
under CA rule 14, which can be easily verified by direct computation.

The last case, formula for $\card \f^{-n}(000)$, follows from the fact that
\begin{equation}
\sum_{b \in \B_3} \card \f^{-n}(b) =\card \B_{2n+3} = 2^{2n+3}.
\end{equation}
This yields
\begin{equation}
\card \f^{-n} (000)=2^{2n+3} - \sum_{i=1}^7 \card \f^{-n}(\beta_i)=(4n+3)C_n.
\end{equation}

\section{Proofs of Lemmas}\label{lemmaproofs}
\subsection{Proof of Lemma~\ref{l0}}
Using definitions of $f_{195}$, $f_{14}$, and $f_{184}$, we obtain after simplification
and after using $x^2=x$ for $x \in \{0,1\}$ the following expressions:
\begin{align} \label{194-14}
 f_{195}(f_{14}&(x_0,x_1,x_2),f_{14}(x_1,x_2,x_3),f_{14}(x_2,x_3,x_4))= \nonumber \\ 
&1- x_1+ x_1  x_0  x_2- x_2  x_1  x_3- x_0  x_2+ x_1  x_0+ x_2  x_3+ x_1  x_3- x_3
\end{align}
and
\begin{align} \label{184-195}
 f_{184}(f_{195}(x_0,x_1,x_2),&f_{195}(x_1,x_2,x_3),f_{195}(x_2,x_3,x_4))=\nonumber \\ 
 &1+ x_1  x_0- x_0  x_2+ x_2  x_3+ x_1  x_3- x_3- x_1.
\end{align}
Subtracting the right hand side of eq (\ref{184-195}) from the right hand side of eq (\ref{194-14}) we obtain $x_1 x_0 x_2-x_2 x_1 x_3$, which vanishes if
one of $x_1$, $x_2$ is equal to zero.
$\square$
 
\subsection{Proof of Lemma~\ref{l1}}
We need to prove that the mapping $M_n$ is indeed a bijection 
from $A:=\mathbf{f}^{-n}_{184}(000)$ to $B:=\mathbf{f}^{-n}_{14}(101)$.
It will be enough to show that $M_n$ is injective, that is, it has the left inverse,
and that $M_n$ is onto.

Before we proceed, let us make two observations. First, note that $\f_{14}^{-1}(101)=\{01001,01010,$ $
01011\}$ and $\f_{14}^{-1}(010)=\{10100,010101,10110,10111\}$. This can be verified by direct computations. As an immediate result, we see
that all elements of $\f_{14}^{-n}(101)$ begin with 101 when n is even, and with 010 when n is odd.

Secondly, again by direct computation, note that $\f_{184}^{-1}(000)=\{00000,00001,00010\}$,
which implies that  elements of $\f_{148}^{-1}(000)$ always begin with 000, for any $n \in \NN$.

For $m \in \NN$, define the block transformation $T(x_0x_1 \ldots x_{m-1})=y_0y_1 \ldots y_{m-1}$
such that 
\begin{equation}
  [T(x)]_i = \begin{cases} 
    0 & \text{if $i=0$},\\    
1+x_{i-1}+x_{i} \mod 2 & \text{ if $i>0$}.
  \end{cases}
\end{equation}
We claim that $T$ is the required left inverse of $M_n$, 
that is, $(T \circ M )(x)=x$ for every $x \in A$. By definition of $T_n$ we have
\begin{equation} \label{TMbothicases}
  [T(M(x))]_i = \begin{cases} 
    0 & \text{if $i=0$},\\    
1+[M(x)]_{i-1}+[M(x)]_{i} \mod 2 & \text{ if $i>0$}.
  \end{cases}
\end{equation}
Since, as observed at the beginning of this proof,  all preimages of 000 start with 0, 
we know that $x_0=0$ for all $x\in A$,
and (\ref{TMbothicases}) yields $[T(M(x))]_i=x_i$ for $i=0$, as required. 
We need to deal with $i>0$ case separately. By definition of $M_n$, for $i>0$
\begin{equation} \label{eqjnonzero}
  [T(M(x))]_i = 1+n+i-1+1+\sum_{l=0}^{i-1} x_l +n + i + 1 + \sum_{l=0}^i x_l \mod 2.
\end{equation}
Since $a+a \mod 2=0$ for any $a\in \{0,1\}$, eq. (\ref{eqjnonzero}) reduces to $[T(M(x))]_i=x_i$,
as necessary for $T$ to be the left inverse of $M_n$. $M_n$ is therefore injective.

To show that $M_n$ is onto, we need to prove that for every $b\in B$ there exists $a$ such that
$a\in A$ and $M_n(a)=b$. Let us choose some $b\in B$ and take $a=T(b)$. 
Using Lemma~\ref{l0}, the fact that 
\begin{equation}
 1+x_{0}+x_{1} \mod 2=f_{195}(x_0,x_1,x_2),
\end{equation}
and knowing that $b$ starts with either 101 or 010,  we can easily show that
\begin{equation}
 T\circ \f_{14}(b)=\f_{184} \circ T(b).
\end{equation}
Therefore, $\f_{184}(T(b))=T(\f_{14}(b)=T(101)=000$, which shows that $a=T(b)$ is indeed a member of  $A$. 
Now we need to show  that $M_n(a)=b$, or equivalently that $M_n(T(b))=b$. By definition of $M_n$,
\begin{equation}
 [M_n(T(b))]_j=n+j+1+\sum_{i=0}^j b_i \mod 2,
\end{equation}
hence
\begin{equation}
  [M_n(T(b))]_j = \begin{cases} 
    n+1 \mod 2 & \text{if $j=0$},\\    
n+j+1+\sum_{i=1}^j (1+b_{i-1}+b_i) \mod 2 & \text{ if $j>0$}.
  \end{cases}
\end{equation}
Basic properties of $\mod 2$ operation reduce this to
\begin{equation} \label{mt}
  [M_n(T(b))]_j = \begin{cases} 
    n+1 \mod 2 & \text{if $j=0$},\\    
n+1+b_0 + b_j \mod 2 & \text{ if $j>0$}.
  \end{cases}
\end{equation}
As remarked at the beginning, 
preimages of 101 begin with 010, while preimages of 010 begin with 101. 
This means that $n$-step preimages of 101 under $\f_{14}$ start either with 0 for odd $n$
or with 1 for even $n$, or equivalently, for $b \in \mathbf{f}^{-n}_{14}(101)$, $b_0=n+1 \mod 2$.
Using this fact, we can see that eq. (\ref{mt}) reduces to  $[T(M(b))]_j=b_j$, which proves
that $M_n(a)=b$ and concludes the proof of surjectivity of $M_n$. 
 $\square$

\subsection{Proof of Lemma~\ref{l3}}
Let $n\in \NN$, and let $x\in \{0,1\}^\ZZ$ be a periodic binary bisequence of period $T\in \NN$ which has the property that all possible blocks $b$
of length $2n+2$ appear in one period of $x$ exactly once. We say that a block $b$ 
of length $m$ \textit{appears} in one period of $x$ if there exists integer $i$, $0\leq i < T$, such that $b_k=x_{i+k}$ for all $k=0,1,\ldots,m-1$.

Existence of  $x$ with the aforementioned property is guaranteed by the existence of a Hamiltonian cycle in de Bruijn graph of dimension $2n+1$
\cite{DeBruij46,Ralston82}. Note that the total number of blocks of length $2n+2$ in $x$ is $2^{2n+2}=4^{n+1}$.

Let us now iterate rule $14$ $n$ times starting with $x$, obtaining $x^\prime$. To be more precise, if
$F_{14}$ is the global function corresponding to $f_{14}$ defined in eq. (\ref{r14def}), we take
$x^\prime = F^n_{14}(x)$.

Since rule 14 conserves the number of blocks $10$
\cite{paper23}, the number of blocks $10$ in one period of $x^\prime$ is the same as in
one period of  $x$. 

Among all possible blocks of
length $2n+2$, exactly $\card \f_{14}^{-n}(10)$ produced block $10$ in one period of $x^\prime$. Thus, the number of blocks 10 in one period of $x^\prime$ is $\card \f_{14}^{-n}(10)$. Because the number of blocks 10 is conserved,
the number of blocks 10 in one period of $x$ is also  $\card \f_{14}^{-n}(10)$. Yet among all possible blocks of length $2n+1$, exactly $1/4$ of all of them begin with 10 (the remaining ones begin with 01, 00, or 11).
Thus, $\card \f_{14}^{-n}(10)=\displaystyle \frac{1}{4} 4^{n+1}=4^n$. Proof for the block $01$ is similar.
 $\square$
\subsection{Proof of Lemma~\ref{l4}}
Let $b \in \f_{14}^{-(n+1)}(010)$. Then $\f_{14}^n(b) \in \f_{14}^{-1}(010)=
\{10100,10101,10110,10111\}$,
and therefore $\f_{14}^n(R^2(b))=101$. This implies $R^2(b)\in \f_{14}^{-n}(101)$,
and therefore 
\begin{equation} \label{inclusion1}
R^2\left(\f_{14}^{-(n+1)}\right) \subset \f_{14}^{-n}(101).
 \end{equation}
Now take a block $a \in \f_{14}^{-n}(101)$, which implies that $\f_{14}^n(a)=101$. For any $u,v\in \{0,1\}$
there exist $u^{\prime},v^{\prime} \in \{0,1\}$ such that
$\f_{14}^n(auv)=101u^{\prime}v^{\prime}$. This implies that
$\f_{14}^{n+1}(auv)=010$, and therefore $auv \in \f_{14}^{-(n+1)}(010)$,
so that $a \in R^2(\f_{14}^{-(n+1)}(010))$. 
This demonstrates that
$\f_{14}^{-n}(101) \subset  R^2\left(\f_{14}^{-(n+1)}\right)$, which together with
eq. (\ref{inclusion1}) leads to the conclusion that sets
$\f_{14}^{-n}(101)$  and $R^2\left(\f_{14}^{-(n+1)}\right)$ are equal, as
required. 
$\square$

\section{Evolution of measures}
We will now come back to the question posed in the introduction.
The appropriate mathematical description of an initial distribution of
configurations is a probability measure $\mu$ on $\calS$. 
Such a measure can be formally constructed as follows. 
If $b$ is a block of length $k$, i.e.,
$b=b_0b_1\ldots b_{k-1}$, then for $i \in \ZZ$ we define a cylinder set
\begin{equation*}
C_i(b)=\{s\in {\cal S}:{ }s_i=b_0, s_{i+1}=b_1 \ldots, s_{i+k-1}=b_{k-1}\}.
\end{equation*}
The cylinder set is thus a set of all possible configurations with fixed
values at a finite number of sites.  Intuitively, measure of the
cylinder  set given by the block $b=b_0\ldots b_{k-1}$, denoted by
 $\mu[C_i(b)]$, is simply a probability of occurrence
of the block $b$ starting at $i$. If the measure $\mu$ is shift-invariant, than
$\mu(C_i(b))$ is independent of $i$, and we will therefore drop the
index $i$ and write simply  $\mu(C(b))$.

The Kolmogorov consistency theorem
states that every probability measure $\mu$ satisfying the consistency
condition
\begin{equation*}
\mu[C_i(b_1\ldots b_k)]=\mu[C_i(b_1\ldots b_k,0)]+\mu[C_i(b_1\ldots b_k,1)]
\end{equation*}
extends to a shift invariant measure on $\cal S$. For $p\in[0,1]$, the Bernoulli
measure defined as $\mu_p[C(b)]=p^j(1-p)^{k-j}$, where $j$ is a number
of ones in $b$ and $k-j$ is a number of zeros in $b$, is an example of
such a shift-invariant (or spatially homogeneous) measure. It describes a set of random
configurations with the probability that a given site is in state $1$
equal to $p$.

Since a cellular automaton rule with global function $F$ maps a configuration in $\calS$
to another configuration in $\calS$, we can define the action of $F$ on measures
on $\calS$.
For all measurable subsets $E$ of $\calS$ we define
$(F\mu)(E)=\mu(F^{-1}(E))$, where $F^{-1}(E)$ is an inverse image of
$E$ under $F$.

If the initial configuration was specified by $\mu_p$, what can be
said about $F^n\mu_p$ (i.e., what is the probability measure after $n$
iterations of $F$)? In particular, given a block $b$, what is the
probability of the occurrence of this block in a configuration obtained
from a random configuration after $t$ iterations of a given rule?

In the simplest case, when $b=1$, we will define the density of ones
as
\begin{equation}
 \rho_n= (F^n\mu_p) (C(1)).
\end{equation}
In what follows, we will assume that the initially measure is symmetric
Bernoulli measure $\mu_{1/2}$, so that the initial density of ones
is $\rho_0=1/2$. 

Assume now that for a given block $b$, the set of $n$-step preimages
is $\f_{14}^{-n}(b)$. Then by the definition of the action of $F_{14}$
on the initial measure, we have
\begin{equation}
 \mu \left(F_{14}^{-n}(C(b)) \right) =  (F^n\mu_p)(C(b)),
\end{equation}
and consequently
\begin{equation}
\sum_{a \in \f_{14}^{-n}(b)}  \mu (a) =  (F^n\mu_p)(C(b)).
\end{equation}

If $b=1$, and $\mu = \mu_{1/2}$, the above reduces to
\begin{equation}
\sum_{a \in \f_{14}^{-n}(1)}  \mu_{1/2} (C(a)) =  \rho_n.
\end{equation}
Note that all blocks $a \in \f_{14}^{-n}(1)$ have length $2n+1$,
and under the symmetric Bernoulli measure, the probability of
their occurrence (that is, $\mu_{1/2}(C(a))$) is the same for
all of them, and equal to $1/2^{2n+1}$. This is because we have
$2^{2n+1}$ of all possible blocks, and each of them is equally probable, so 
a single one has probability $1/2^{2n+1}$. As a result, we obtain
\begin{equation}
 \rho_n=2^{-2n-1}\card  \f_{14}^{-n}(1).
\end{equation}
In rule 14, preimages of $1$ are 001, 010, and 011. We can therefore
write
\begin{equation}
 \rho_n=2^{-2n-1}\left(\card  \f_{14}^{-(n-1)}(001)+\f_{14}^{-(n-1)}(010)+\f_{14}^{-(n-1)}(011)\right).
\end{equation}
Using formulas of Theorem \ref{maintheorem}, this yields
\begin{equation}
 \rho_n=2^{-2n-1}\left( 4^{n} - (2 n-1) C_{n-1} \right)
=\frac{1}{2}\left(1-\frac{2 n-1}{4^n}C_{n-1} \right).
\end{equation}
Using Stirling's formula to approximate factorials in the definition
of $C_{n-1}$, after elementary calculations one obtains asymptotic
approximation valid for $n \to \infty$,
\begin{equation}
 \rho_n \approx \frac{1}{2} -\frac{1}{4 \sqrt{\pi}}n^{-\frac{1}{2}}.
\end{equation}

Very similar calculations can be performed for other block
probabilities. Defining the probability of occurrence
of a block $b$ in a configuration obtained from the random initial
configuration after $n$ iterations by $P_n(b)=(F^n\mu_{1/2})(C(b))$
we obtain 
\begin{equation}
 P_n(b)=\sum_{a \in \f_{14}^{-n}(b)}  \mu_{1/2} (C(a)).
\end{equation}
If the length of block $b$ is denoted by $|b|$, its $n$-step
preimage has length $|b|+2n$. Again, since all blocks are equally
probable in $\mu_{1/2}$,  a single block of length $|b|+2n$
has probability $2^{-|b|-2n}$, and we obtain
\begin{equation}
 P_n(b) = 2^{-|b|-2n} \card \f_{14}^{-n}(b).
\end{equation}
Since cardinalities of blocks of length up to 3 are known,
the above result together with formulas of Theorem \ref{maintheorem}
can be used to derive exact probabilities of blocks of length up to 3. 

\section{Conlcusions}
In closing, let us remark that rule 14 belongs to class 3 rules
in informal Wolfram's classification. Its dynamics, while 
not as complicated as other class 3 rules, is far from simple.
It is therefore encouraging to find meaningful regularities in 
preimage sets of this rule, and to be able to compute probabilities
of small-length cylinder sets exactly, without any approximations.

 One hopes that systematic analysis
of properties of preimage trees of other elementary rules
reveals more regularities of this type, perhaps leading to
more general results. To speculate a bit, one may conjecture
that rules having additive invariants (like rule 14) are
likely to possess regularities in their preimage trees.
Further investigation of this issue is currently ongoing.


\providecommand{\href}[2]{#2}\begingroup\raggedright\endgroup

\end{document}